\documentclass[aps,pra,superscriptaddress,twocolumn,10pt,longbibliography]{revtex4-2}

\usepackage{amsmath}
\usepackage{graphicx}
\usepackage{bm}
\usepackage{color}

\date{\today}

\newcommand{\sz}{S^z}
\newcommand{\sx}{S^x}
\newcommand{\sy}{S^y}

\definecolor{darkblue}{rgb}{0.1,0.2,0.6}
\definecolor{darkred}{rgb}{0.8,0.1,0.2}
\usepackage[colorlinks,citecolor=darkblue,linkcolor=darkred,urlcolor=darkblue]
{hyperref}
\usepackage[all]{hypcap}

\usepackage{etoolbox}
\usepackage{mathtools}
\usepackage{amsmath}%
\usepackage{amsfonts}%
\usepackage{amssymb}

\begin{document}

\title{Stark many-body localization: Evidence for Hilbert-space shattering}
\author{Elmer~V.~H.~Doggen}
\email[Corresponding author: ]{elmer.doggen@kit.edu}
\affiliation{\mbox{Institute for Quantum Materials and Technologies, Karlsruhe Institute of Technology, 76021 Karlsruhe, Germany}}
\author{Igor~V.~Gornyi}
\affiliation{\mbox{Institute for Quantum Materials and Technologies, Karlsruhe Institute of Technology, 76021 Karlsruhe, Germany}}
\affiliation{\mbox{Institut f\"ur Theorie der Kondensierten Materie, Karlsruhe Institute of Technology, 76128 Karlsruhe, Germany}}
\affiliation{Ioffe Institute, 194021 St. Petersburg, Russia}
\author{Dmitry~G.~Polyakov}
\affiliation{\mbox{Institute for Quantum Materials and Technologies, Karlsruhe Institute of Technology, 76021 Karlsruhe, Germany}}

\begin{abstract}
We study the dynamics of an interacting quantum spin chain under the application of a linearly increasing field. This model exhibits a type of localization known as Stark many-body localization. The dynamics shows a strong dependence on the initial conditions, indicating that the system violates the conventional (``strong'') eigenstate thermalization hypothesis at any finite gradient of the field. This is contrary to reports of a numerically observed ergodic phase. Therefore, the localization is crucially distinct from disorder-driven many-body localization, in agreement with recent predictions on the basis of localization via Hilbert-space shattering.
\end{abstract}

\maketitle

\emph{Introduction.}---
Sufficiently strong disorder can localize an interacting many-body system at finite energy density of excitations (even at high temperature), a phenomenon known as many-body localization (MBL) \cite{Gornyi2005a, Basko2006a, Nandkishore2015a, Altman2015a, Abanin2017a, Alet2018a}. This can be thought of as a generalization of Anderson localization \cite{Anderson1958a, Evers2008a} to the many-body case. Signatures of the MBL phase were found experimentally, where it was observed that the dynamics of disordered systems can halt on laboratory time scales \cite{Kondov2015a, Schreiber2015a, Choi2016a, Chiaro2019a}.

Recently, it was shown by Schulz \textit{et al.}~\cite{Schulz2019a} and van Nieuwenburg \textit{et al.}~\cite{vanNieuwenburg2019a} that some features of the MBL phase can appear in systems without disorder. The key ingredient is an applied field with (approximately) constant gradient. From a single-particle perspective, this applied field induces Wannier-Stark localization. It was then demonstrated, building on earlier works using exact numerical results for small chains \cite{Tomadin2007a} and the nonlinear Schrödinger equation \cite{GarciaMata2009a}, that this localization is robust to the introduction of interactions for a sufficiently strong applied field, a phenomenon termed Stark (or Bloch) many-body localization (SMBL). The authors furthermore argued that a transition from ergodicity to localization should emerge at some finite value of the applied field. However, the authors employed exact diagonalization for relatively small chains up to $L = 24$ sites. It is still debated whether such studies provide reliable insight into the conceptually more interesting thermodynamic limit \cite{Khemani2017a, Panda2019a, Abanin2019a, Sierant2019a}. It was shown numerically that the MBL transition for the archetypal Heisenberg chain with on-site disorder shifts substantially with system size \cite{Doggen2018a}.

As a means to explain the observation of SMBL, recent theoretical studies propose \emph{Hilbert-space shattering} \cite{Khemani2020a} or \emph{Hilbert-space fragmentation} \cite{Sala2020a} (see also Refs.~\cite{Taylor2020a, Herviou2020a}). The key idea is that the Hilbert space of the system fragments into disconnected sectors, thus preventing thermalization. While the main focus is on fractonic models \cite{Nandkishore2019a}, the authors of Ref.~\cite{Khemani2020a} report a general proof that is applicable to a wide variety of systems exhibiting local constraints, and they argue that this also hinders thermalization in SMBL systems, at least for a certain class of initial states at large values of the applied field \footnote{The relation between fractonic models and the model under consideration here is further discussed in Ref.~\cite{Taylor2020a}.}. It is argued that at large values of the field gradient, localization might persist up to exponentially long (or even infinite) timescales. This can be regarded as an analogy with the formation of many-body quantum scars \cite{Turner2018a}.

From the perspective of experiments, a recent study reports the observation of SMBL in a superconducting quantum processor \cite{Guo2020a}. SMBL has also been investigated in cold-atom setups in a two-dimensional system tilted in one direction \cite{Guardado2020a}, and more recently in a one-dimensional system \cite{Scherg2020a}.

\begin{figure}
    \centering
    \includegraphics[width=\columnwidth]{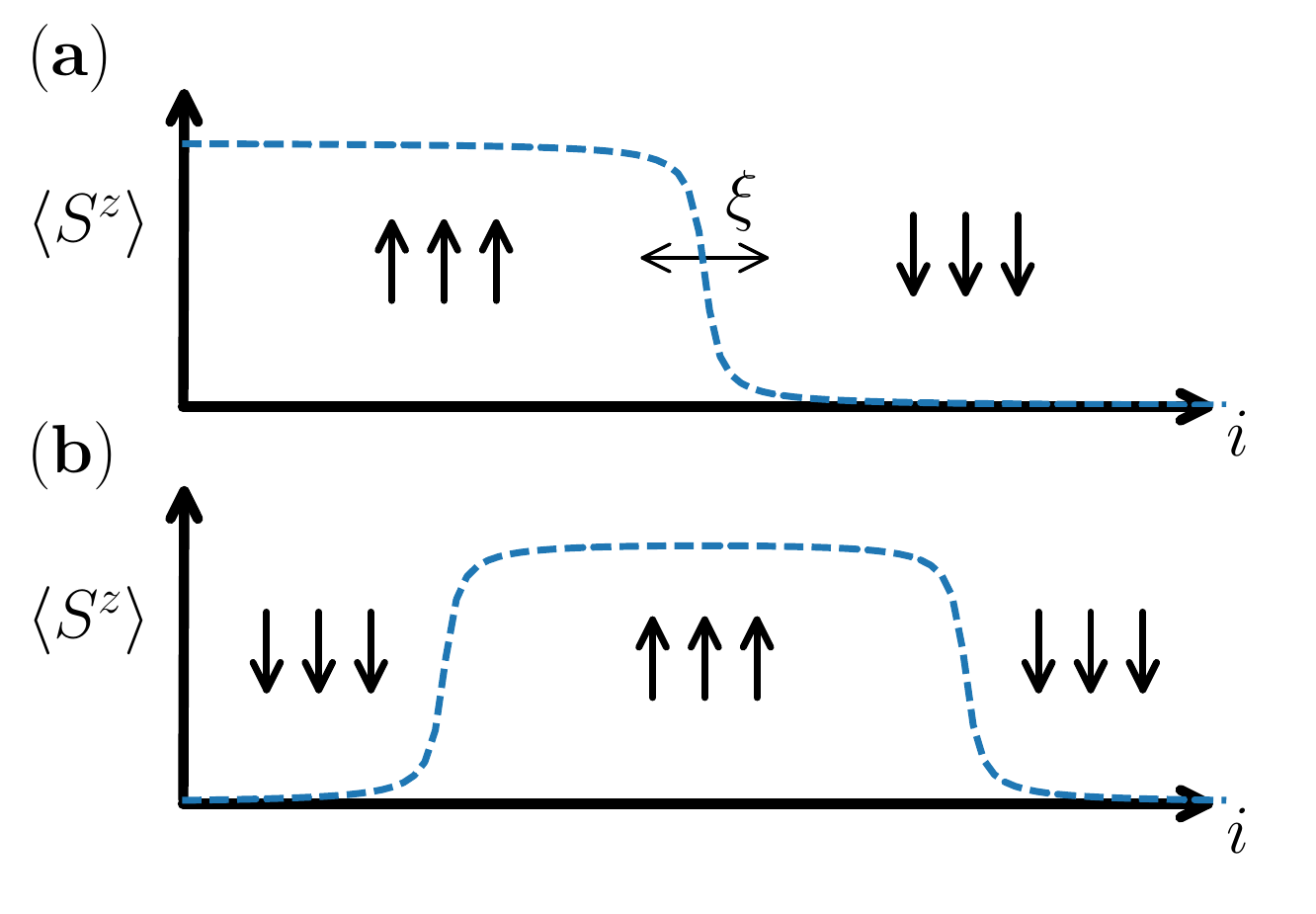}
    \caption{Cartoon of the stationary state of the spin density $\langle S^z \rangle$ as a function of position $i$ after initializing with (a) a single domain wall and (b) two domain walls. In the former case, particle transport beyond a region of characteristic width $\xi$ is prohibited by global energy conservation; in the latter case by local energy conservation.}
    \label{fig:cartoon}
\end{figure}

It is therefore of interest to numerically investigate the fate of one-dimensional SMBL as reported in Refs.~\cite{Schulz2019a, vanNieuwenburg2019a, Wu2019a, Moudgalya2019a, Taylor2020a, Yao2020a, Zhang2020a, Klockner2020a, Guo2020a, Scherg2020a, Lerose2020a}. For this purpose we employ the powerful numerical machinery of the time-dependent variational principle (TDVP) \cite{Haegeman2016a}, a method formulated in terms of the language of matrix product states (MPS). The TDVP allows for investigation of system sizes far beyond those available to exact diagonalization, at the cost of being limited to short times or weakly ergodic systems. Since MBL systems are weakly entangled, this makes the TDVP especially powerful as a tool for numerical analysis of such systems \cite{Kloss2018a, Doggen2018a, Doggen2019a, Chanda2020a, Doggen2021a}.

Here we show, on the basis of numerical results and semiclassical considerations, that the aforementioned Hilbert-space shattering precludes thermalization (in a sense defined below) in the model exhibiting SMBL. These results indicate that, in the thermodynamic limit, a certain class of initial states exhibits nonergodicity at least up to exponentially long times, for both weak and strong gradients.

\emph{Model.}--- We consider a one-dimensional spin chain on a lattice of length $L$ with open boundary conditions, as described by the Hamiltonian
\begin{equation}
 \mathcal{H} = \sum_{i=1}^L \epsilon_i \sz_i + \sum_{i=1}^{L-1} \Big[ J(\sx_i\sx_{i+1} + \sy_i \sy_{i+1}) + \Delta \sz_i \sz_{i+1} \Big]. \label{eq:hamiltonian}
\end{equation}
Here $\sx$,  $\sy$, and $\sz$ are spin-$1/2$ operators. An on-site field is applied with strength $\epsilon_i = Wi$ (we take $W > 0$ w.l.o.g.). In the limit $\Delta \rightarrow 0$, this problem can be mapped onto non-interacting spinless fermions using a Jordan-Wigner transformation. In that case, it is well-known that the eigenfunctions of the Hamiltonian \eqref{eq:hamiltonian} are localized due to Wannier-Stark localization.  In the following, we take $J = \Delta = 1$.
In the case of no applied field $W = 0$, the model is the isotropic Heisenberg chain, which thermalizes rapidly \cite{DAlessio2016a}.

\emph{Absence of an ETH-MBL transition at finite gradient.}---
Here we argue that the model \eqref{eq:hamiltonian} does not permit a transition from an ergodic phase to a localized one at finite $W$, assuming that the ergodic phase satisfies the eigenstate thermalization hypothesis (ETH) \cite{DAlessio2016a}. Here we mean the ETH in a ``strong'' sense, meaning that the long-time behavior of local observables does not depend on the initial state for \emph{any} choice of initial state, as long as it has the same relevant macroscopic quantities. For a closed system with fixed particle number (microcanonical ensemble), the quantity of interest is the energy. We will comment below on the possible implications for the ``weak'' ETH, which requires only that \emph{most} initial states (possibly excluding some states with measure zero in the Hilbert space) exhibit the aforementioned loss of memory.

Consider the scenario depicted in Fig.~\ref{fig:cartoon}, where the system of length $L$ is initialized with a domain wall in the middle of the system, such that $\langle S^z \rangle = 0$. The dynamics of the system is bound by conservation of energy and particle or spin transport to the right boundary of the system is, just like in the non-interacting case, inhibited. No matter the value of $W$, there will be a certain finite length scale $\xi \equiv \xi(W)$ that characterizes the width of the domain wall in the late-time limit. In the thermodynamic limit the system is then localized for any $W > 0$, where the thermodynamic limit is understood as the limit of taking $L \rightarrow \infty$, followed by $t \rightarrow \infty$. However, this is clearly only a specific choice of the energy and not a general case.

\begin{figure*}
    \centering
    \includegraphics[width=\textwidth]{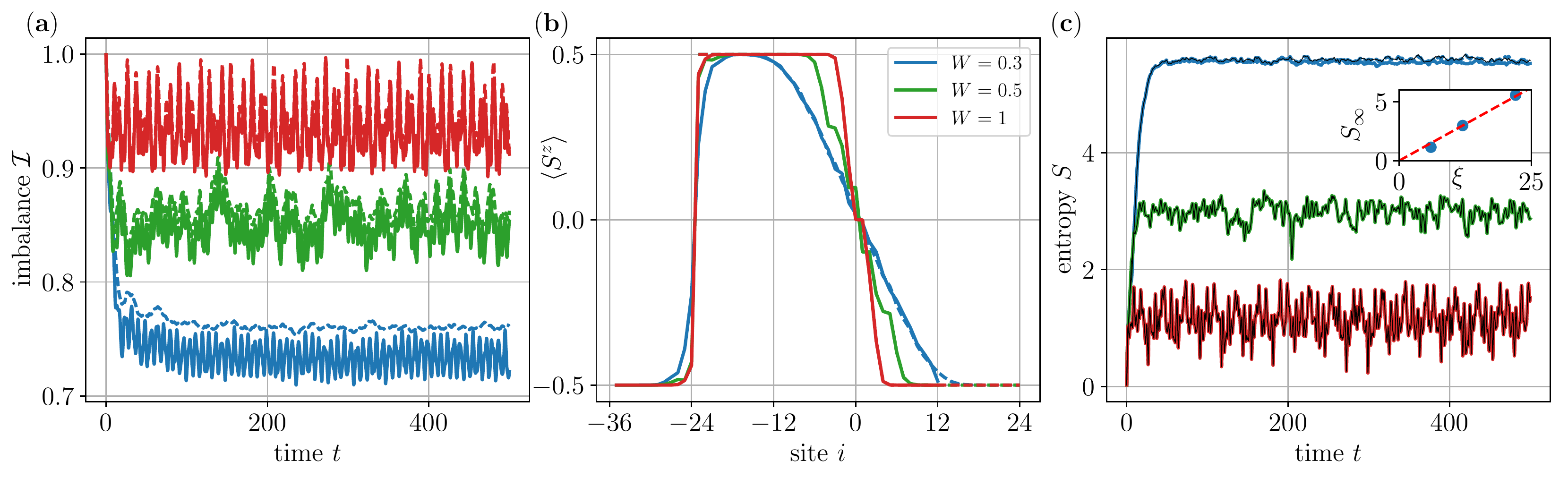}
    \caption{Dynamics for the double-domain wall initial state (solid lines) and the single domain wall initial state (dashed lines) for $L=48, \Delta = 1$. (a) Imbalance dynamics \eqref{eq:imba} as a function of time for various choices of $W$. (b) Spin density $\langle S^z \rangle$ at $t=500$ for both initial conditions, with an offset such that the rightmost domain wall is at $i=0$. (c) Von Neumann entropy of entanglement, with the bipartition at the rightmost domain wall [$i=0$ in panel (b)]. Thin black lines indicate $S$ for the single domain wall and $L = 32$. Inset: saturated entropy $S_\infty$, defined as the average entropy over the time interval $[100, 500]$, as a function of $\xi$, defined as the extent of the region around $i = 0$ with $|\langle S^z (t=500) \rangle| < 0.45$. The red dashed line is a linear fit through the origin.}
    \label{fig:imb}
\end{figure*}

Nonetheless, the above argument can be generalized to arbitrary energy by introducing more domain walls. Consider the initial state shown in Fig.~\ref{fig:cartoon}, with two domain walls. If those walls are situated at roughly $L/4$ and $3L/4$, then we again have a state with $\langle S^z \rangle = 0$, but now with an energy that is close to the center of the energy band. With a fixed distance $L/2$ between the domain walls a desired energy can be reached (up to arbitrary precision in the thermodynamic limit) by shifting the initial position of the walls. The dynamics, however, is locally identical to the first scenario, while the global energy constraint no longer prevents delocalization. Thermalization of the ETH-type would require transport processes that link both domain walls. However, those processes are exponentially suppressed in the thermodynamic limit since then the requirement $\xi \ll L$ is always satisfied. In other words, the Hilbert space is shattered or fragmented \cite{Khemani2020a, Sala2020a}, and the two-wall initial state, or more generally any state with sufficiently large polarized regions, can persist at least up to exponentially large timescales in $\xi/L$.

If, instead, we consider a finite \emph{density} of domain walls (i.e., a charge density wave initial condition), say $\eta$, then we can identify two regimes: i) $\eta \xi \ll 1$: the system is localized in the sense discussed above, since the different domain walls are far enough apart such that the reasoning above applies; ii) $\eta \xi \gg 1$: the system is ``quasi-ergodic'' and may appear thermal in the sense of exhibiting ergodic grains of finite size. If one considers a finite-size system, then such grains can exceed the size of the system. While the system in this case is not thermalizing according to the ETH, it will still show ETH-like level spacing statistics, given sufficient integrability-breaking. This is an essential difference with the non-interacting case, where no ETH-like level statistics are observed.

In a strict sense, the ETH requires that the Hilbert space is not shattered, yet the reasoning above implies that such shattering occurs at any $W > 0$. Exponentially slow processes might allow for large polarized regions to melt eventually. Faster thermalization has been predicted on the basis of resonances in the single-particle spectrum in the case where the field is purely linear \cite{Taylor2020a}. Those processes might be responsible for local thermalization within the region $\xi$. As we shall see, however, the numerical results provide no indication of full thermalization for $\xi \eta \lesssim 1$ on laboratory timescales even for a constant field gradient.

It is instructive to compare the scenario considered above to the ``standard'' MBL problem of the isotropic Heisenberg chain with on-site disorder \cite{Luitz2015a}. In that case, the choice of initial state is not important for the long-time state of the system. Compared to a N\'eel initial condition $|\uparrow \downarrow \ldots \rangle$, the domain wall initial condition merely thermalizes somewhat more slowly in the ergodic phase \cite{Hauschild2016a}, and there is no analogous constraint related to local or global energy conservation. The arguments outlined above do not apply for standard MBL.

\emph{Method.}---
We compute the dynamics starting from a product state initial condition in the $S^z$-basis with $\sum_i \langle S^z_i \rangle = 0$. By tracking the density imbalance
\begin{equation}
    \mathcal{I}(t) = \frac{4}{L} \sum_i \langle S^z_i (t) \rangle \langle S^z_i (t=0)\rangle , \label{eq:imba}
\end{equation}
one can conveniently determine whether the time-evolved state has a memory of the initial state, in the sense of the ETH.

Our numerical method is the time-dependent variational principle (TDVP) as applied to matrix product states (MPS) \cite{Haegeman2016a}. We use a similar method as in previous works \cite{Doggen2018a, Doggen2019a, Doggen2020a}; 
 the reader is referred to Ref.~\cite{Paeckel2019a} for a recent review of MPS-based methods to simulate dynamics. The method boils down to the projection of the unitary dynamics onto the variational manifold of the MPS:

\begin{equation}
 \frac{d}{dt} |\psi \rangle = -i \mathcal{P}_\mathrm{MPS}\mathcal{H}|\psi\rangle. \label{eq:tdvp}
\end{equation}

We use a bond dimension $\chi = 256$ (which determines the size of the variational manifold) and compute dynamics up to $t=500$. Details are provided in the Supplementary Material \cite{SupMat}.

\begin{figure*}
    \centering
    \includegraphics[width=\textwidth]{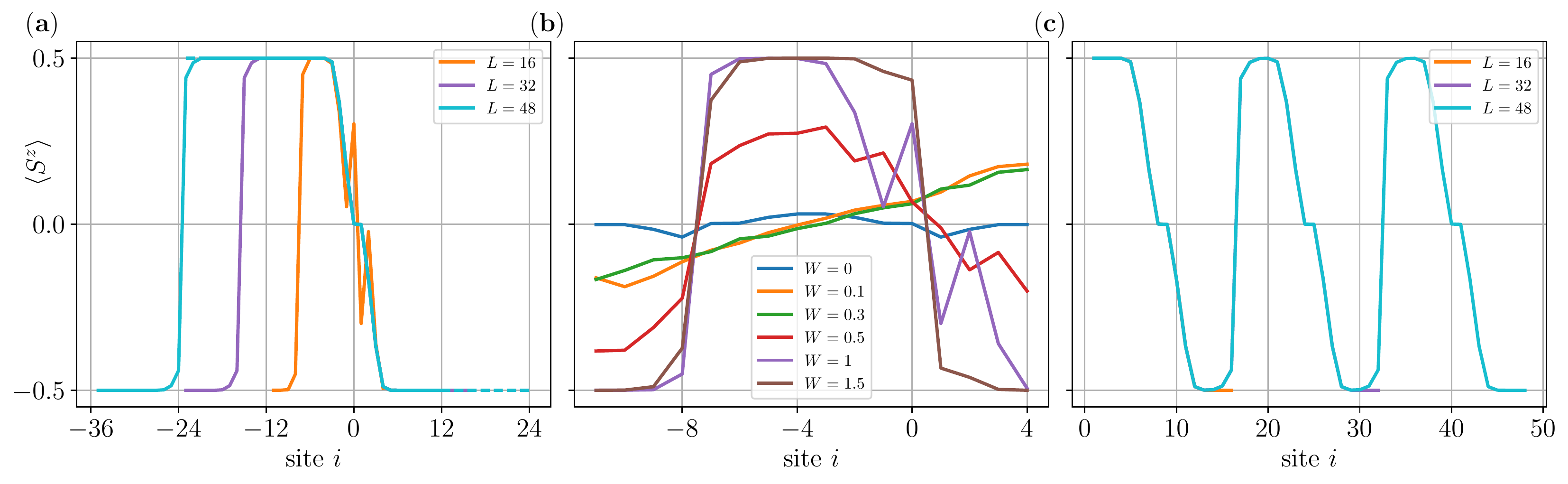}
    \caption{(a) spin density $\langle S^z \rangle$ at $t=500$ for various system sizes $L = 16, 32, 48$ and fixed $W = 1$ and both the double domain wall (solid lines) and single domain wall (dashed lines) initial conditions, cf.~Fig.~\ref{fig:imb}b. The difference between system sizes, visible in the ``zig-zag'' shape for $L = 16$, is due to boundary effects. (b) $\langle S^z \rangle$ at $t=500$ for $L=16$ and various $W$, using the double domain wall initial condition. (c) as in panel (a), but for a fixed domain wall density $\eta = 1/8$. }
    \label{fig:diffL}
\end{figure*}

\emph{Density and imbalance dynamics.}---
We consider the Hamiltonian \eqref{eq:hamiltonian} with open boundary conditions. We can compare the cases where the initial state is a double domain wall to a single domain wall, cf.~Fig.~\ref{fig:cartoon}. In Fig.~\ref{fig:imb}a we show the imbalance dynamics. The domain wall melting quickly halts, even for a relatively weak value of the field gradient $W = 0.3$. The late-time spin density is shown in Fig.~\ref{fig:imb}b. While the global energies of the initial states are very different, the local dynamics is practically identical, showing the system does not thermalize up to the times considered here. Moreover, the dynamics does not depend strongly on system size as long as $\xi \lesssim 1/\eta$, see Fig.~\ref{fig:diffL}a. As $W$ is decreased, at a certain value dependent on the size of the domains, the melting domains meet, see Fig.~\ref{fig:diffL}b. Note the difference between the results for $W = 0.3, L = 16$ and $W = 0.3, L = 48$: the former is clearly delocalized, while the latter is localized. Indeed, upon inspection of Fig.~\ref{fig:imb}b, we can infer that $\xi(W=1) \approx 8$, which corresponds to the crossover value $W = 1$ observed for $L = 16$ with $\eta = 1/8$. Hence, $W = 0.3$ and $\eta = 1/8$ corresponds to the ``quasi-ergodic'' regime discussed above.

There is a significant difference between the dynamics at either side of the domain wall. This can be explained by considering the relative signs of the interaction $\Delta$ and the field $W$. If they have the same sign (as in the case we consider here), then it is easier for spin waves to move ``uphill.'' This is because spins will lose energy by moving apart due to the antiferromagnetic coupling, which is then compensated by the increased potential energy from the linear field. However, at the left domain wall, spins can only move ``downhill'': spins lose energy by the domain wall melting as well as through reduced antiferromagnetic coupling and very little melting of the domain wall is permitted by local energy conservation. This phenomenon was dubbed ``negative current'' in Ref.~\cite{Klockner2020a} and is another sign of non-thermalization.

\emph{Entropy.}---
A frequently used measure for the ETH-MBL transition in disordered systems is the bipartite von Neumann entropy of entanglement:

\begin{equation}
    S \ln 2 = -\mathrm{Tr}\Big[ \rho_A \ln \rho_A \Big], \quad \rho_A = \mathrm{Tr}_\mathrm{B} \rho,
\end{equation}
where the system is divided into two segments $A$ and $B$, and $\rho$ is the density matrix. In Fig.~\ref{fig:imb}c we show the results for the entanglement entropy, where the division between segments is at the rightmost domain wall $\cdots \uparrow \uparrow \downarrow \downarrow \cdots$ for both configurations. No logarithmic growth of the entanglement entropy with time is observed, but instead a fast crossover from linear growth to saturation at a constant value $S_\infty$ -- a hallmark of the localized phase. There is a very weak dependence on system size: the entropy curves are plotted for different system sizes $L = 32$ (single wall) and $L = 48$ (double wall), which indicates saturation with system size and hence area-law scaling of the entropy. At the same time we can see very little difference between the single and double-domain wall configurations. The similarity between the different initial conditions -- which have a strongly different global energy -- is another direct signature of the suppression of entanglement across the entire system, since it implies that only the sites close to the position of the initial domain wall contribute to the entropy. Indeed, the asymptotic value of the entropy $S_\infty$ scales linearly with $\xi(W)$ (see the inset of Fig.~\ref{fig:imb}c). Note that $S_\infty$ determines the smallest $W$ that can be reached numerically, since the numerical $\mathrm{max}(S) = \log_2(\chi)$. For $\xi \ll L$, we find this leads, with reasonable numerical resources, to $W \gtrsim 0.3$ (see Supplementary Material \cite{SupMat}). This value is deep into the regime identified as ergodic in Refs.~\cite{Schulz2019a,vanNieuwenburg2019a}, where it is argued that a transition from an ergodic to a localized regime occurs at $W \approx 1$.

\begin{figure}
    \centering
    \includegraphics[width=.95\columnwidth]{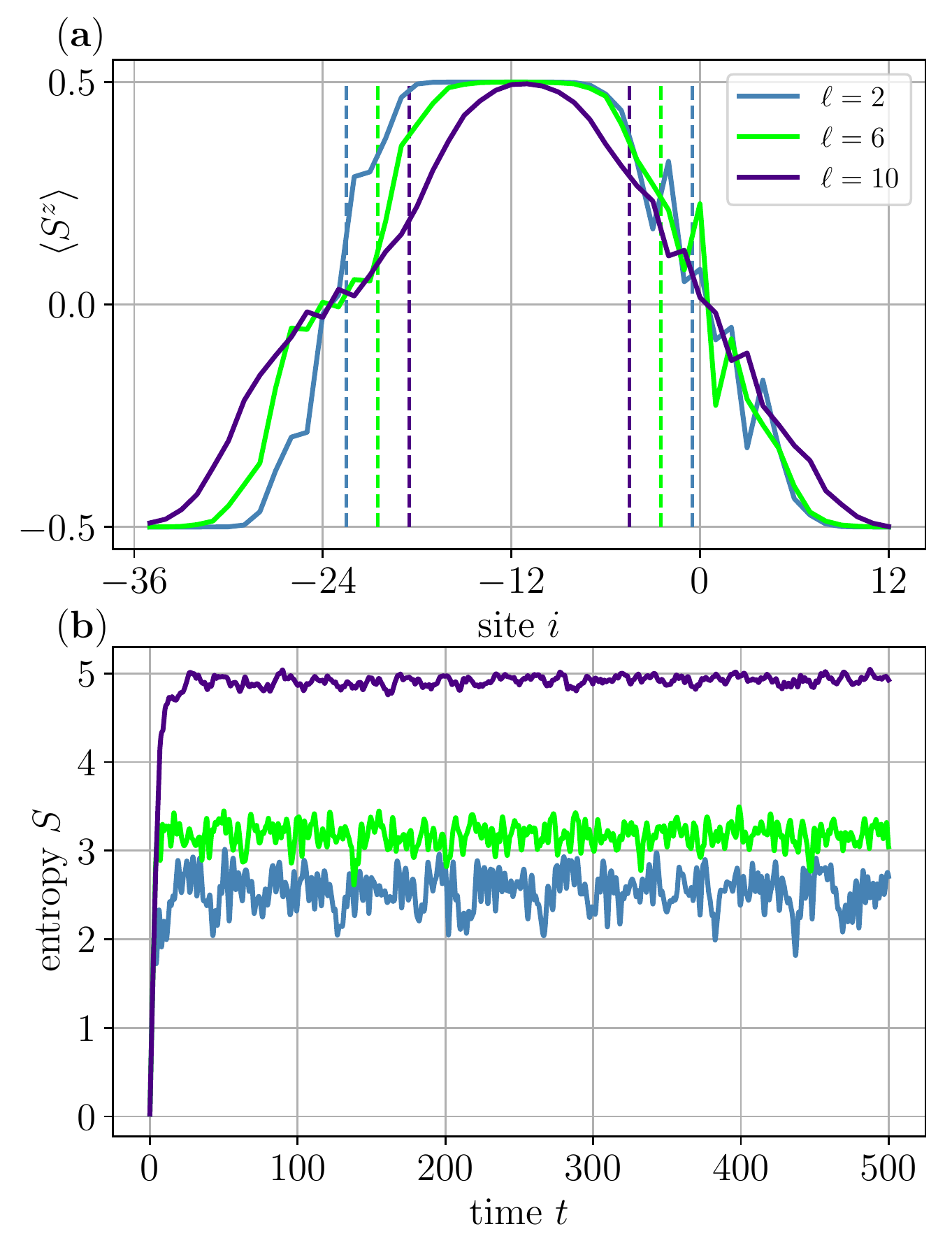}
    \caption{(a) $\langle S^z \rangle$ at $t=500$ for $L=48$ and $W = 0.5$, where the initial condition contains N\'eel regions of width $\ell$ (see main text). Dashed vertical lines indicate the extent of the polarized region in the middle at $t = 0$, varying with $\ell$. (b) Von Neumann entropy of entanglement, with the bipartition at the center of the rightmost N\'eel region.}
    \label{fig:Neel}
\end{figure}

\emph{Stability of the domain wall.}---
Thus far, we have considered two types of initial conditions: a single domain wall and a double domain wall configuration. One may wonder how robust the halting of transport and the growth of entanglement is to changing the initial condition. We consider the following adaptation: instead of a ``hard'' domain wall in the double-domain wall configuration, we replace the step in spin density by a N\'eel region of length $\ell$. For example, by flipping two spins we obtain an initial state $\cdots \uparrow \uparrow \downarrow \uparrow \downarrow \downarrow \cdots$, corresponding to a N\'eel region of length $\ell = 2$. The introduction of this region acts as a region of effectively high local temperature, and might be expected to aid delocalization. Numerical results for the spin and entanglement dynamics are shown in Fig.~\ref{fig:Neel} for the choice $W = 0.5$. Comparing Figs.~\ref{fig:Neel}a and \ref{fig:imb}b, we observe that the melting of the polarized region (vertical dashed lines in Fig.~\ref{fig:Neel}b) at the right domain wall is largely unaffected by introducing the N\'eel region, whereas the left density profile looks more similar to the right one. Despite this ``symmetrization,'' a finite polarized region remains in the long-time limit. Both the halting of transport and the saturation of the entropy of entanglement are therefore robust upon including these N\'eel regions to the system. Local thermalizing regions do not lead to melting across larger distances and fine-tuning the initial condition is not necessary for observing non-ergodicity.

\emph{Discussion.}--- We have investigated the fate of Stark many-body localization. Crucial differences between this type of localization and disorder-driven many-body localization are elucidated. In particular, we show that there is no ergodic phase described by the strong ETH in such systems, as the late-time dynamics is strongly dependent on initial conditions, in agreement with the predicted mechanism of Hilbert-space shattering \cite{Khemani2020a, Sala2020a}. As can be seen from the dynamics, $\xi \approx 8$ for $W \approx 1$, so if one chooses an initial state with $\eta = 1$ (the N\'eel state) then $W \approx 1$ is already deep into the ``quasi-ergodic'' regime where melted regions overlap strongly. This explains why an apparent transition is observed numerically \cite{Schulz2019a,vanNieuwenburg2019a}, yet it is not of the strong ETH-type since the thermalization (in the quasi-ergodic sense) depends strongly on the initial state, and for large systems one can always find initial states, such as the double-wall configuration, that would fully thermalize at most exponentially slowly in $\xi \eta$, if at all. Quasi-ergodic features manifest themselves only in local thermalization within the melted region. The dependence on the initial state (through $\eta$ or an equivalent quantity) furthermore prohibits the definition of a mobility edge dependent only on energy. Such a mobility edge was reported recently \cite{Zhang2020a}. In contrast, we find that states with strongly different energy (the single and double domain wall configurations) can have locally identical dynamics up to long times.

While these results convincingly show the strong ETH is violated, the fate of the weak ETH remains an open question. Nevertheless, from the investigation of the dynamics for the double-domain wall configuration with N\'eel regions we can infer that the critical feature leading to the halting of transport and entanglement growth (at least up to the timescales considered here) is only the length of the largest polarized region of the system. In the space of all possible random product states, we will find with probability one a sufficiently long polarized region of length $\lambda$ with $\xi \ll \lambda \ll L$ at $L \rightarrow \infty$, so that such a state will show similar dynamics. Moreover, there will be exponentially many (in $L$) such states, which confirms the ``shattering'' of the Hilbert space into a large number of disconnected sectors.

Experimentally, the system considered here was studied in Refs.~\cite{Guo2020a,Scherg2020a}. The authors of Ref.~\cite{Scherg2020a} report a rather robust non-ergodic behavior, consistent with our predictions here. The thermalizing phase observed in Ref.~\cite{Guo2020a} can be identified with the quasi-ergodic regime in our work, i.e., as a finite-size effect.

It is worth pointing out more differences between disorder-driven MBL and SMBL. In the former case, there is substantial evidence that the transition point from the ETH-type phase to the localized phase is itself localized, as per the ``avalanche'' mechanism \cite{DeRoeck2017a,Thiery2017a, Dumitrescu2019a, Goremykina2019a, Morningstar2020a, Doggen2020a}. There is no evidence of such a localized transition point for the Wannier-Stark system. Indeed, there is no quenched disorder present, and no subdiffusion is observed. Moreover, as emphasized above, we do not observe logarithmic growth of the entanglement in time \cite{Bardarson2012a}.

In conclusion, care must be exercised when drawing parallels between SMBL and ``conventional'' MBL. In the former case, the localization is due to Hilbert space shattering, while in the latter case due to emergent local integrals of motion.

\emph{Note added.}--- After the initial submission of this manuscript a related preprint appeared \cite{Yao2021a}, confirming the non-ergodic properties observed here in the case of large polarized regions. In addition, an experimental study appeared \cite{Morong2021a},  confirming the existence of the Stark-MBL phase in a system with interacting trapped ions and observing a nonzero imbalance even with modest values of the field gradient.

\emph{Acknowledgments.---} We are grateful to D.~A.~Huse for interesting discussions, in particular for suggesting to investigate the initial condition considered in Fig.~\ref{fig:Neel}. Numerical simulations were performed using the TeNPy library (version 0.6.1) \cite{tenpy}. I.V.G.~acknowledges support from Deutsche Forschungsgemeinschaft (DFG) Grant No. GO 1405/6-1 and the Russian Foundation for Basic Research (RFBR), Grant No. 18-02-01016.

\bibliography{ref}

\end{document}


\title{Supplementary Material to ``Stark many-body localization: Evidence for Hilbert-space shattering''}

\maketitle

\renewcommand{\theequation}{S\arabic{equation}}
\renewcommand{\thefigure}{S\arabic{figure}}
\setcounter{equation}{0}
\setcounter{figure}{0}

\section{Numerical details}

As described in the main text, the time evolution is obtained by means of the time-dependent variational principle (TDVP), a matrix product state method. Such methods rely on a variational ansatz for the many-body wave function, with a number of variational parameters controlled by the bond dimension $\chi$ \cite{Schollwock2011a}. As we have argued, we can define a characteristic size of the melted region of a domain wall, $\xi \equiv \xi(W)$, and a density of domain walls in the initial condition $\eta$. In the limit of $\xi \eta \ll 1$, we should expect that distant regions in are completely disconnected and hence exhibit no entanglement. We therefore would expect the low-entanglement ansatz of the MPS to work well.

\begin{figure*}[!h]
    \centering
    \includegraphics[width=\columnwidth]{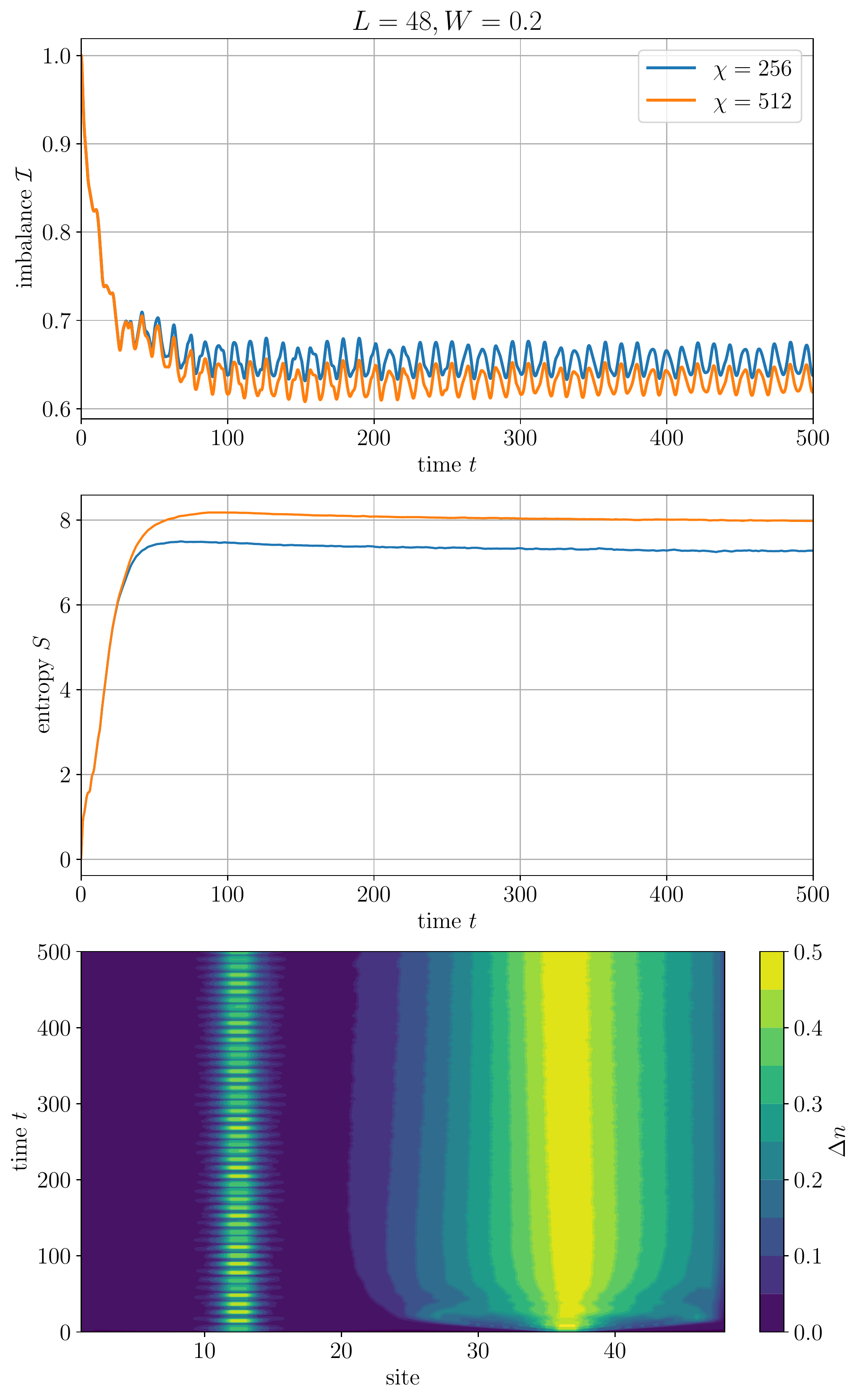}
    \includegraphics[width=\columnwidth]{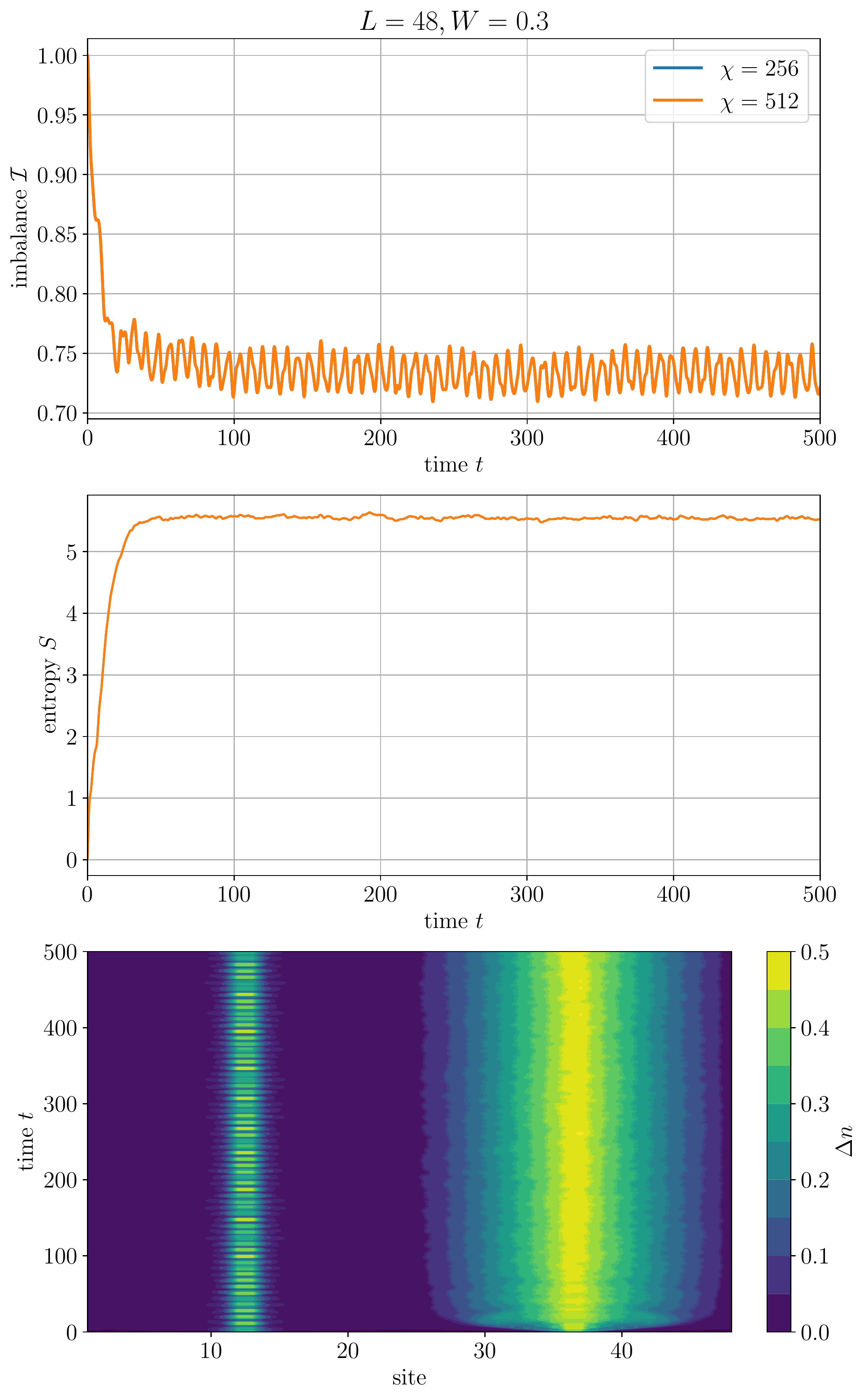}
    \caption{Dynamics for the double domain wall initial condition, for $L = 48$, $\eta = 1/24$ and the slope of the potential $W = 0.2$ (left) and $W = 0.3$ (right). Top panel: the imbalance as a function of time for two choices of bond dimension $\chi = 256$ and $\chi = 512$. Middle panel: entropy as a function of time for the same choices of $\chi$, where the bipartition is taken at the rightmost domain wall. Bottom panel: difference of the density at individual sites in the system with respect to the initial density $\Delta n$ as a function of time for $\chi = 512$.}
    \label{fig:bench02}
\end{figure*}

Consider Fig.~2b of the main text. For $W = 0.3$, we can estimate the size of the melted region as $\xi \approx 24$ sites, and the density of domain walls in the chosen initial condition is $\eta = 1/24$ (two domain walls with a system size $L = 48$). We therefore expect numerical convergence will be lost around $W = 0.3$. This is precisely what is observed, as shown in the figure \ref{fig:bench02} where results are shown for $W = 0.2$ and $W = 0.3$, with bond dimensions $\chi = 256$ and $\chi = 512$.

\bibliography{ref}